\documentclass[11pt, leqno]{article}
\usepackage{times, amsfonts, amsmath, amssymb, latexsym, setspace, epsfig}
\usepackage{subfigure, graphics}

\def\bs{\boldsymbol}

\def\Pr{{\rm I\!P}}
\def\be{\begin{equation}}
\def\ee{\end{equation}}
\def\bea{\begin{eqnarray*}}
\def\eea{\end{eqnarray*}}
\def\bean{\begin{eqnarray}}
\def\eean{\end{eqnarray}}

\def\ra{\rightarrow}

\def\Bl{\Bigl}
\def\Br{\Bigr}

\def\R{{\bf R}}

\def\alp{\alpha}

\def\eps{\epsilon}

\begin{document}
\setstretch{1.1}

\title{Calibrating the scan statistic with size-dependent critical values: heuristics, methodology and computation}

\author{Guenther Walther\thanks{Work supported by NSF grants DMS-1501767 and DMS-1916074} \\
        Department of Statistics, 390 Serra Mall \\
        Stanford University, Stanford, CA 94305 \\
        gwalther@stanford.edu}
\date{February 2022}
\maketitle

\begin{abstract}
It is known that the scan statistic with variable window size favors the detection of signals with small
spatial extent and there is a corresponding loss of power for signals with large spatial extent. Recent results have shown
that this loss is not inevitable: Using critical values that depend on the size of the window allows optimal
detection for all signal sizes simultaneously, so there is no substantial price to pay for not knowing the correct window
size and for scanning with a variable window size.
This paper gives a review of the heuristics and methodology for such size-dependent critical values, their
applications to various settings including the multivariate case, and recent results about fast algorithms
for computing scan statistics.
\end{abstract}

\vfill
\vfill

\noindent\textbf{Keywords and phrases.} Scan statistic, scale-dependent critical values, multiscale inference,
blocked scan, Bonferroni scan, fast algorithm.

\noindent\textbf{MSC 2000 subject classifications.} Primary 62G10; secondary 62G32.

\newpage

\section{Introduction}  \label{introduction}

The scan statistic is the standard tool for detecting a signal that
may be present at an unknown location in a sequence of observations or in a random field. 
There has been a considerable recent growth in the popularity of the scan statistic due to
the emergence of related problems that involve the detection of anomalies in e.g. sensor
arrays and digital images, see Arias-Castro et al.~(2011) for further examples.
It turns out that several key methodological issues that arise in these diverse settings  are already present
in a simple univariate sequence model and that the solutions derived there can often be readily transferred
to these more involved settings. It is therefore helpful
to focus on the simple univariate model where we observe
\be \label{model}
Y_i\ =\ \mu {\bs 1}(j\leq i \leq k) +Z_i,\ \ i=1,\ldots,n
\ee
where the $Z_i$ are i.i.d. N(0,1) and both the mean $\mu$ and the support $[j,k]$ are
unknown. The task is to decide whether an elevated mean is present, i.e. to test $\mu=0$
vs. $\mu >0$ when the support $[j,k]$ is unknown.

The scan statistic arises in this setting using a standard statistical approach as follows:
The log-likelihood ratio statistic for testing $\mu=0$ when the support $[j,k]$ is known is computed
to be $\frac{1}{2} (\sum_{i=j}^k Y_i)^2/(k-j+1)$. Since $[j,k]$ is unknown, maximizing over $j$ and $k$
yields the generalized likelihood ratio test  statistic
\be  \label{scan}
M_n\ =\ \max_{1\leq j \leq k \leq n} \frac{\sum_{i=j}^k Y_i}{\sqrt{k-j+1}},
\ee
which is the scan statistic. It should be noted that the scan statistic was originally defined
for a fixed window width $k-j$, see Glaz et al.~(2001), but as pointed out in Cressie (1977), Loader (1991),
Nagarwalla~(1996) and Kulldorff~(1997), 
the window width is typically not known and therefore also needs to be
maximized over\footnote{In recognition of this historical precedent Sharpnack and Arias-Castro~(2016)
call $M_n$ the multiscale scan, but this paper will simply refer to $M_n$ as the scan.}. 
Note that treating the window width as unknown would seemingly greatly increase
the multiple testing penalty of the problem as in addition to scanning over the unknown location $j$
one also has to scan over the unknown width $k-j$.
It is a rather recent and perhaps surprising finding that scanning over an
unknown window width can be done in a way that does not result in a substantial loss of detection
power when compared to the case where the correct window width is known a priori. In other words,
while there is a price to pay for scanning over locations, there is no further substantial
penalty for scanning over the unknown window width, provided that the scan is calibrated in
an appropriate way. This review paper describes different methods for such a calibration and
gives some heuristic understanding of the theoretical results in the literature that motivate
such an approach.

Scan statistics appear in different stochastic settings. 
In fact, one of the first problems that motivated the use of the scan statistic
was the detection of time intervals with an unusually large number of events, see Glaz et al.~(2001).
 While the technical
work required in the latter setting is different from that in the Gaussian model (\ref{model}), it turns
out that the heuristics and general ideas described below apply just as well,
and theoretical results
are remarkably similar in these different settings. 
Section~\ref{other} describes how these heuristics apply to
some of these situations. Computing the scan statistic becomes a critical issue
in the multivariate setting, and the past decade has seen important work in that regard which is
summarized in Section~\ref{computation}. One key
result is that computing certain multivariate scan statistics is not subject to the curse of dimensionality.
That is, there are algorithms that compute a sufficiently precise approximation to the scan statistic
with a time complexity that is almost linear in the number of observations.
This unexpected result has the potential to shift the direction of future statistical work on the scan
statistic, which in the last 50 years was focused on deriving its exact or approximate null distribution:
Fast algorithms make it possible to efficiently simulate
this null distribution, thus moving the focus of research to constructing such algorithms for the particular
setting at hand, and to other methodological problems such as how to combine the evidence from
various scan windows in a way different from (\ref{scan}). In fact, it is another interesting finding
emerging in the last decade that good solutions to the computational and statistical problems
often rest on the same key ideas, as will be seen below.

\section{Heuristics for calibrating the maximum}

While the scan statistic (\ref{scan}) enjoys a simple form and is motivated as generalized likelihood
ratio test, it was observed that it is very sensitive for very small window
sizes at the expense of moderate and larger window sizes, see Naus and Wallenstein~(2004), Neill~(2009)
and Chan~(2009). Naus and Wallenstein (2004)  propose to remedy this situation
by first finding the cluster size (window width) $w$ that results in the smallest p-value $p_w$
(computed for this single $w$), and then taking as overall p-value the probability
that any cluster of smaller or equal size will have a p-value smaller than $p_w$ under the null
hypothesis. A detailed investigation of the performance of this method is still outstanding.
This exposition will focus on different ways to address the calibration problem that have
become increasingly popular in the last decade.

In order to understand the problem at hand it is helpful to rewrite the scan (\ref{scan})
as 
\be \label{scan2}
 M_n \ =\ \max_{1\leq w \leq n} L_w,
\ee
where $L_w=\max_{1 \leq j \leq n+1-w} \sum_{i=j}^{j+w-1} Y_i/\sqrt{w}$ is the maximum over all
windows with width $w$. The standard use of the scan rejects $H_0:\,\mu=0$ if $M_n$ exceeds its critical
value $q_n(\alp)$, which is defined under $H_0$ via
\be \label{calibtrad}
\Pr_0 \Bl(M_n > q_n(\alp)\Br)\ =\ \alp
\ee
for a given significance level $\alp \in (0,1)$. Note that this implies that all $L_w$ are referred
to the same critical value $q_n(\alp)$. But it is readily seen that one can as well use critical
values that depend on the window width $w$: Any vector $\{q_{n,w}(\alp), w=1,\ldots,n\}$
that satisfies 
\be \label{sizedepend}
\Pr_0 \Bl(L_w > q_{n,w}(\alp)\ \mbox{ for some } w \in \{1,\ldots,n\}\Br)\ \leq \ \alp
\ee
will result in a test that has level $\alp$. The motivation for using such size-dependent
critical values derives from a desire to counteract the abovementioned observation that $M_n$ 
is very sensitive to small window sizes at the expense of moderate and larger window sizes.
While it is clear from (\ref{sizedepend}) that size-dependent critical values allow to trade off
the sensitivity of various window sizes, it would also appear from (\ref{sizedepend}) that such
a trade-off is a zero-sum game: Lowering $q_{n,w}(\alp)$ (and therefore increasing the power
of $L_{w}$) for some $w$ necessarily requires increasing some of the other $q_{n,w}(\alp)$ in order to
keep the significance level $\alp$, therefore reducing the
power of those $L_w$. Surprisingly, it is possible to construct $q_{n,w}(\alp)$ such that this
trade-off is not a zero-sum game but nearly a `free lunch': it is possible to substantially
increase the power for moderate and large window sizes at the cost of a very small (and asymptotically
disappearing) power loss for small window sizes. Before describing relevant methodology we give a heuristic 
explanation of the underlying idea, which exploits the concentration of Gaussian maxima:

Consider the smallest window size $w=1$. Then $L_w$ is the maximum of $n$ i.i.d. standard
normals, and this maximum is known to be very close to $\sqrt{2 \log n}$, see e.g. Leadbetter et al.~(1983).
An analogous conclusion holds for larger window sizes $w>1$: since there are about $\frac{n}{w}$ abutting
and hence disjoint windows, there are about $\frac{n}{w}$ independent standard normals in $L_w$. Their
maximum is close to $\sqrt{2 \log \frac{n}{w}}$. This remains true for the maximum over all (overlapping)
intervals of length $w$ because the corresponding test statistics are strongly correlated with the
$\frac{n}{w}$ independent ones. Therefore the distribution of (\ref{scan2}) is dominated by small
windows with $w\approx 1$ and concentrates around $\sqrt{2 \log n}$. 
This  explains the observation by Naus and Wallenstein~(2004)
that the scan is sensitive for very small windows at the expense of moderate and larger window
sizes: The null distribution of $L_w$ is close to $\sqrt{2 \log \frac{n}{w}}$, but the null 
distribution of $M_n$ will concentrate around $\sqrt{2 \log n}$, thus resulting 
in a power loss for larger windows. 

But the above heuristic also suggests a way to remedy this problem: It suggests that one can increase
$L_w$ by as much as $\sqrt{2 \log n}- \sqrt{2 \log \frac{n}{w}}$ without increasing the maximum
$M_n\approx \sqrt{2 \log n}$, provided that $L_w$ is concentrated closely around $\sqrt{2 \log \frac{n}{w}}$. 
The consequence of this would be an increase in power for larger window sizes $w$; in
fact, the resulting critical value for $L_w$ would be $\sqrt{2 \log \frac{n}{w}}$, which is
the same critical value we would obtain for $L_w$ if we tested the window size $w$ only. In other words,
there would be no multiple testing penalty for considering multiple window sizes $w$ in the thus
modified $M_n$. 

This heuristic suggests to calibrate the scan via
\be \label{DS}
\widetilde{M}_n\ :=\ \max_{1\leq w \leq n} \Bl( L_w -\sqrt{2 \log \frac{n}{w}}\Br).
\ee
This results in size-dependent critical values for $L_w$ of the form
\be \label{DScalib}
q_{n,w}(\alp)\ =\ \sqrt{2 \log \frac{n}{w}} +\kappa_n(\alp)
\ee
where $\kappa_n(\alp)$ denotes the $(1-\alp)$-quantile of $\widetilde{M}_n$. This choice 
ensures  that (\ref{sizedepend}) holds, i.e. the resulting test has level $\alp$.
The penalty term $\sqrt{2 \log \frac{n}{w}}$ appears to have been first introduced by
D\"{u}mbgen and Spokoiny~(2001) in the context of function estimation.

\subsection{Asymptotic optimality theory and finite sample performance} \label{finiteopt}

The heuristic derivation of the solution (\ref{DScalib}) to the problem
(\ref{sizedepend}) of finding size-dependent critical values can be backed up with a theoretical
optimality result. This result was established by D\"{u}mbgen and Spokoiny~(2001) in the different 
context of testing certain features of a function. In the context of the scan problem (\ref{model}),
this result states that the calibration (\ref{DScalib}) results in asymptotic
power 1 against alternatives 
\be \label{optimal}
\mu = \mu (n,w) \geq (\sqrt{2} +\eps_n) \sqrt{\frac{\log \frac{n}{w}}{w}},
\ee
provided that $\eps_n$ is not too small: $ \eps_n \sqrt{\log \frac{n}{w}} \ra \infty$. On
the other hand, they show that if `$\sqrt{2} +\eps_n$' is replaced by `$\sqrt{2} -\eps_n$',
then no test can exist that will detect this alternative with nontrivial asymptotic power in
the minimax sense. Thus there is a sharp cutoff below which detection is not possible, while
above the cutoff $\widetilde{M}_n$ has asymptotic power 1 and is therefore optimal.
It can be shown that these thresholds also apply if the window size $w$ is known a priori,
in which case no multiple testing across various $w$ is required. Therefore this result formalizes
the above heuristic finding that the calibration (\ref{DScalib}) avoids the multiple testing
penalty for considering multiple window sizes $w$, at least in a large sample setting.
This above sharp cutoff applies to small window sizes, i.e. $\frac{n}{w} \ra \infty$. A complementary
optimality result for larger window sizes, i.e. $\limsup \frac{n}{w} < \infty$, shows that the above
condition (\ref{optimal}) is necessary for asymptotic power 1, see D\"{u}mbgen and Walther~(2008) and 
Datta and Sen~(2018). Note that in that case (\ref{optimal}) becomes $\mu \sqrt{n} \ra \infty$, which is
the parametric rate for detecting an elevated interval, as one would expect.

In contrast, it was shown in Chan and Walther~(2013) that the regular scan statistic $M_n$
in (\ref{scan}) will not attain these optimal thresholds except for the very smallest window
sizes.


The theoretical optimality results described above concern the large sample setting.
However, simulations show that the finite sample performance of the calibration
(\ref{DS},\ref{DScalib}) is often unsatisfactory as it is very sensitive for large windows at the expense
of very small windows, i.e. it overcompensates the shortcoming of the scan $M_n$, see Rufibach
and Walther~(2010), Sharpnack and Arias-Castro~(2016) and Walther and Perry~(2019).
The reason for this overcompensation is a consequence of Gaussian concentration around
the maximum: $L_w=\sqrt{2 \log \frac{n}{w}} +o_p(1)$, where the $o_p(1)$ term 
becomes small as $\frac{n}{w} \ra \infty$. So
for larger window sizes the $o_p(1)$ term need not be small, and in fact will not be small.
Therefore, the null distribution of $\widetilde{M}_n$ will essentially be determined
by the large windows, with a resulting loss of power for very small windows that is not adequately
captured by the asymptotic optimality theory. See Walther and Perry~(2019)
for simulations that detail this effect. They also explain why the asymptotic
optimality theory described above will necessarily be too imprecise to evaluate the performance of a calibration
with an appropriate level of prevision, even in the large sample context. In summary, the asymptotic
optimality theory available at the time of this writing provides a condition that is necessary but not sufficient
for a good performance.

In lieu of evaluating the performance of a calibration via asymptotic optimality, Walther and Perry~(2019)
propose a finite sample criterion: the {\sl realized exponent} measures how close the calibration comes
to attaining the asymptotic detection boundary in the finite sample situation at hand.
This is especially relevant in a multivariate situation, where asymptotic results for the massive
multiple testing problem involved in scanning appear to be a poor indication of the performance
in a finite sample situation, see Section~\ref{multivariate}.

\section{Methodology for calibrating the scan statistic}

This section describes methodology that results in improved power for larger windows vis-a-vis the scan
(\ref{scan2}) as motivated by
the above heuristics, while largely avoiding the overcompensation (i.e.loss of power at small $w$) of (\ref{DS}).

\begin{enumerate}
{\bf \item The Sharpnack-Arias-Castro calibration}

Walther and Perry~(2019) report that a modification of a penalty introduced by Sharpnack and
Arias-Castro~(2016), namely
$$
SAC_n\ =\ \max_{1\leq w \leq n} \Bl( L_w -\sqrt{2 \log \Bl[\frac{en}{w}(1+\log w)^2\Br]}\Br)
$$
gives the desired improvement to (\ref{DS}): It results in better detection power for larger windows than the
scan (\ref{scan2}), while sacrificing  only a slight amount of power at small windows.
As in the case (\ref{DS}), the Sharpnack-Arias-Castro calibration results in size-dependent critical values
for $L_w$, which are now of the form
\be  \label{SACcalib}
q_{n,w}(\alp)\ =\ \sqrt{2 \log \Bl[\frac{en}{w}(1+\log w)^2\Br]} +\kappa_n(\alp)
\ee
where $\kappa_n(\alp)$ now denotes the $(1-\alp)$-quantile of $SAC_n$. 
This quantile $\kappa_n(\alp)$
can be obtained by simulation, resulting in finite sample confidence statements. 

The Sharpnack-Arias-Castro calibration is closely related to the D\"{u}mbgen-Spokoiny calibration (\ref{DS}),
but in the former case the penalty term is larger for larger window sizes, thus assigning more power to
small window sizes. In fact, D\"{u}mbgen and Spokoiny~(2001) give a refinement of their penalty that aims
for the same outcome by introducing an iterated logarithm term. However, Walther and Perry~(2019) report
that this refinement has a performance that is nearly indistinguishable from (\ref{DS}) for sample sizes
up to $10^6$.

 Note that the Sharpnack-Arias-Castro calibration
is tailor-made for the Gaussian setting (\ref{model}), not only because of the simulation of the quantile $\kappa_n(\alp)$,
but importantly the penalty term in $SAC_n$ will produce a good distribution of power across the window lengths
$w$ specifically in the (sub-)Gaussian case, without any guarantee for a reasonable performance for other
distributions. If one wishes to use $SAC_n$ for different settings and retain its good performance, then it is 
necessary to employ a test statistic that has sub-Gaussian tails under the null hypothesis, or to transform
the test statistic $\frac{\sum_{i=j}^k Y_i}{\sqrt{k-j+1}}$ to this effect. Fortunately, this is possible
in many important cases. Walther~(2021) gives a number of ways to standardize sums in order to obtain
sub-Gaussian, or even Gaussian, tail bounds,  e.g. by working with the square root of the log likelihood ratio statistic
in the case of exponential families, see also Rivara and Walther~(2013), Frick et al. (2014),
K\"{o}nig et al. (2018); by employing
self-normalized sums in the case of symmetric distributions; or by working with statistics based on ranks
in the case of exchangeable random variables. Of course, constructing a test
statistic that has sub-Gaussian tails under the null is not sufficient for a good performance, the particular
test statistic must also result in good detection power. This has to be established in each setting with
theory or with simulations.

\item {\bf The blocked scan}

All of the preceding solutions to the calibration problem (\ref{sizedepend}) use
an explicit specification of the critical values $q_{n,w}$. A different approach is given
in Walther~(2010), who proposes to calibrate
the significance levels of the various window sizes rather than the critical values.
The idea is that by directly controlling these significance levels
one can ensure a desirable trade-off between small and large windows
in a quite general way that is not specific to the Gaussian setting; recall that the penalty
terms above were derived for the Gaussian setting and different distributional assumptions would
require a different penalty.

The idea of the blocked scan is to group intervals of roughly the same length (e.g.
having length within a factor of 2) into a block. Then all intervals within a block are assigned the same critical
value. This critical value is set so that the significance level for the block is proportional to a harmonic
sequence (i.e. a harmonic sequence in the block index when the blocks are indexed by increasing window size).
We refer the reader to Walther and Perry~(2019) for the details.
Simulations and theory in Walther~(2010) and Walther and Perry~(2019) show that this calibration performs 
similarly to the Sharpnack-Arias-Castro calibration in the Gaussian case and that it possesses asymptotic
optimality properties.

The blocked scan has the advantage that it provides a quite general recipe
that is neither wedded to a distributional assumption nor to the univariate setting, and
which has also proven useful in related testing problems, see e.g. Rufibach and Walther~(2010).
On the other hand, the blocked scan requires to obtain the quantiles for each block by simulation,
and the number of simulation runs will be larger than for e.g. (\ref{DS}) since one needs to simulate
multiple quantiles. This is being facilitated by the development of fast approximations
to the scan statistic, in particular in the multivariate setting, that are described in Section~\ref{computation}.
Another key point for an efficient simulation is the fact that the number of blocks is small, of the order $\log_2 n$, 
due to the grouping of intervals by length within a factor of 2.

An alternative to simulating the quantiles of each block would be to use a model for their distribution.
Results by Siegmund and Venkatraman~(1995), Kabluchko~(2011)
and Sharpnack and Arias-Castro~(2016) in the Gaussian model (\ref{model}) suggest
that the distribution of $\max_{w \in \ell\mbox{th block}} L_w$ can be approximated by
a Gumbel distribution after an appropriate centering and scaling. Then one needs to simulate
only one parameter $\tilde{\alp}$, which is to be used for all blocks,
in order to calibrate the blocked scan to have level $\alp$,
see Walther~(2010) or Walther and Perry~(2019).
Note that the Gumbel distribution would only be used as a model for the block in order to make it possible
to easily balance the contributions of  the blocks (i.e.
window sizes) to the overall statistic. Even if the approximation via the Gumbel distribution turned out to be poor,
the choice of $\tilde{\alp}$ will ensure that the test has a valid level $\alp$.
 
\item {\bf The Bonferroni scan}

Each of the calibrations considered so far, namely (\ref{scan}), (\ref{DS}), $SAC_n$ and the blocked scan,
involves a maximum, and therefore the null distribution needs to be approximated either analytically
or by simulation. The Bonferroni scan avoids this extraneous computation by calibrating the statistic
for each window, i.e. $\sum_{i=j}^k Y_i/\sqrt{k-j+1}$, with a weighted Bonferroni adjustment.
At first glance it would appear that applying a Bonferroni correction to the $\sim n^2$ windows
$(j,k]$ in model (\ref{model}) is ill-advised as it would lead to a hopeless loss of power - even more
so in a $d$-dimensional setting, where the number of rectangles grows like $n^{2d}$.
However, work in the last 15 years has shown that it is possible to effectively approximate the maximum
of these $\sim n^2$ statistics $\sum_{i=j}^k Y_i/\sqrt{k-j+1}$ by evaluating them on a judiciously
chosen set of only about $O(n)$ windows. The Bonferroni scan exploits this fact not only for fast
computation, but also for inference: Since it suffices to consider a sparse collection of relatively
few windows, there is hope that a simple Bonferroni adjustment will lead to critical values
that result in good performance. It was shown in Walther and Perry~(2019) that this is indeed the case:
Using a certain weighted Bonferroni adjustment leads to a performance in terms of detection power that
is asymptotically optimal, and  almost as good as that
of the Sharpnack-Arias-Castro calibration and the blocked scan in a finite sample setting. 
In fact, the mechanics of the weighted Bonferroni
adjustment are very similar to the blocked scan: Windows are grouped into blocks according to their length
and the Bonferroni adjustment includes a weight that is proportional to a harmonic sequence.
The choice of the approximating set of windows is crucial for a good performance of the Bonferroni scan.
This requires to strike a delicate balance in order to guarantee a good detection power: 
the collection of windows has to be rich enough so it can
provide a good approximation for an arbitrary window $(j,k]$, but it also has to be sparse enough so 
that a Bonferroni adjustment will not unduly diminish the detection power.
The main idea is that instead of evaluating all windows $(j,k]$, $0\leq j < k \leq n$, one considers
only endpoints on a grid $\{i d_{\ell},\, i=0,1,2,\ldots \}$, where the spacing $d_{\ell}$ has to
increase with the window length $\ell$ at a certain rate. See Section~\ref{computation} and
Walther and Perry~(2019) for the details.
\end{enumerate}

For all of the above three calibrations, the improved performance compared to (\ref{DS}) is achieved
by relenting somewhat on the demand for optimality for the largest windows $w \approx n$,
which is presumably not an important concern in practice.
For example, the penalty in $SAC_n$ will not stay bounded for large windows
and hence the optimality result for large windows will not hold any more. But for these large
windows the penalty grows only at the very slow rate
$\sqrt{\log \log n}$, which should result in only a small loss of power.

With the exception of the Bonferroni scan, all of the calibrations discussed so far
involve terms that need to be approximated analytically or by simulation, 
such as $q_n(\alp)$ for the traditional scan (\ref{calibtrad}) or
$\kappa_n(\alp)$ for (\ref{DScalib}) or (\ref{SACcalib}).
There exist large sample approximations to the critical values of several of these
calibrations, see Siegmund and Venkatraman~(1995),
Kabluchko~(2011), and  Sharpnack and Arias-Castro~(2016). However, Sharpnack and Arias-Castro~(2016)
find that these approximations are often too conservative, and they recommend instead to approximate
these critical values with simulations or permutation tests. Such simulations and permutation
tests can often be rigorously justified, see e.g. Walther~(2010), Rivera and Walther~(2013), or
Arias-Castro et al.~(2018),
and are made possible by recent advances that allow for a fast computation of the scan statistic, 
see Section~\ref{computation}.

\section{Related settings}
\label{other}

As mentioned in Section~\ref{introduction}, scan statistics often arise in settings that
are different from the Gaussian sequence model (\ref{model}). Of particular interest in
literature is the setting where one observes an (in)homogeneous Poisson process and the problem
is to detect an interval where the intensity is elevated compared to a known baseline, i.e. one
is looking for an interval with an unusually large number of events, see Glaz et al.~(2001).
Conditioning on the total number of observed events allows to eliminate certain nuisance parameters
and shows that the problem is equivalent to testing whether $n$ i.i.d. observations arise from
a known density $f_0$ (which w.l.o.g. can be taken to be the uniform density on $[0,1]$)
 versus the alternative that $f_0$ is elevated on an interval $I$:
\be  \label{density}
f_{r,I}(x)\ =\ \frac{r 1(x \in I) + 1(x \in I^c)}{rF_0(I) +F_0(I^c)}f_0(x),
\ee
so the problem becomes testing $r=1$ vs. $r>1$, see Loader (1991) and Rivera and Walther (2013). 
The results in the latter paper show that the heuristics, methodology and optimality results
in the density/intensity model (\ref{density}) are quite analogous to the Gaussian sequence
model (\ref{model}). However, in order to apply the calibrations from the previous section and
to achieve optimal detection, it is critical to set up the methodology correctly. In particular,
it was pointed out in Rivera and Walther~(2013) that using the {\sl square root} of the log likelihood 
ratio statistic together with the calibrations described in the previous section provides
a general recipe for optimal detection. That is, just as the statistic 
$\frac{\sum_{i=j}^k Y_i}{\sqrt{k-j+1}}$ in (\ref{scan})  corresponds
to the square root of twice the log likelihood ratio statistic in (\ref{model}), one needs
to work with the  square root of twice the log likelihood ratio statistic in (\ref{density})
rather than statistics such as $Y_{(k)}-Y_{(j)}$ which are often used in model (\ref{density})
to detect clusters of points.
The reason why this is important is that the $\sqrt{2 \log \ldots }$ penalty term in calibrations such as
(\ref{DScalib}) or $SAC_n$ derives from the Gaussian tail of the statistic $\sum_{i=j}^k Y_i$. But 
statistics such as $Y_{(k)}-Y_{(j)}$ do not follow
a normal distribution under the null hypothesis, and therefore these calibrations 
will not give satisfactory results. On the other hand, the log likelihood ratio statistic       
provides a quite general solution to this problem: 
Not only is log likelihood ratio known to typically provide the most powerful
test statistic, but according to statistical folklore twice the log likelihood ratio follows
approximately a $\chi_1^2$ distribution, so that one can hope that the square root of twice the log
likelihood ratio will have approximately Gaussian tails and therefore calibrations such as those
discussed above should provide optimal detection. It was indeed shown in Rivera and Walther (2013) that 
the square root of twice the log likelihood ratio statistic in (\ref{density}) satisfies a
finite sample sub-Gaussian tail inequality.
Therefore, the methodology and optimality results from the Gaussian sequence model (\ref{model})
can be carried over to the model (\ref{density}). This transfer of methodology rests on
a second key point: The indices $(j,k)$ for the window in (\ref{scan})
correspond to the indices of order statistics in model (\ref{density}), i.e. the scanning is with respect
to the empirical measure rather than with respect to the sample space. With that identification
the scan statistic becomes a process on $\{1,\ldots,n\}$ rather than on a particular sample space,
and the methodology described in the previous sections can be applied quite generally by
replacing $\frac{\sum_{i=j}^k Y_i}{\sqrt{k-j+1}}$ with $\sqrt{2\, logLR(j,k)}$, where
$logLR(j,k)$ is the log likelihood ratio statistic on the interval $(Y_{(j)}, Y_{(k)})$. 

The square root of the log likelihood ratio statistic has been successfully applied in other
contexts, e.g. in Bernoulli model with a random design in Walther (2010), or to data
from an exponential family observed at a fixed design in Frick et al. (2014) and in
K\"{o}nig et al. (2018).
In the case of exponential families the crucial sub-Gaussian tail bound for the square root of the log
likelihood ratio can be derived by inverting the Cram\'{e}r-Chernoff bound, see Rivera and Walther~(2013)
and Walther~(2021);  or by normal approximation, see Frick et al. (2014) and K\"{o}nig et al. (2018);
or under the permutation distribution, see Walther~(2010).

\section{The multivariate case}  \label{multivariate}

Surprisingly, the heuristics, methodology and optimality results described in the previous
sections also apply in the multivariate setting, with no fundamentally different changes to the respective
formulas if the scanning windows are sufficiently regular, e.g. balls or rectangles (even with variable
aspect ratio). To appreciate this result, note that the heuristic behind the optimal detection
threshold (\ref{optimal}) is the well-known fact that the maximum of $n$ i.i.d. standard
normals concentrates around $\sqrt{2 \log n}$. Now suppose we were to maximize over $n^d$ rather
than $n$ independent normals, i.e. we massively increase the scale of the multiple testing
problem. Since $\sqrt{2 \log n^d} = \sqrt{(2d) \log n}$, one sees that the constant $\sqrt{2}$
in the detection threshold (\ref{optimal}) measures in a certain sense the difficulty of the
detection problem and therefore plays a crucial role. This explains the effort to
bound the constant in (\ref{optimal}) as tightly as possible, in contrast to many other
theoretical results in the statistics literature where constants typically play a minor role.
In fact, due to its importance, Arias-Castro et al.~(2005) call the constant $d$ 
in the threshold $\sqrt{(2d) \log n}$ the {\sl exponent of effective dimension}.

At first sight, such a massive increase in the multiple testing problem 
is to be expected by scanning in $d$-dimensional rather than in univariate space.
For example, there are of the order $\sim n^2$ univariate
intervals that contain distinct subsets of $n$ points sampled from $U[0,1]$
(since there are $\sim n$ left endpoints and
$\sim n$ right endpoints). In contrast, there are $\sim n^{2d}$ axis-parallel rectangles
that contain distinct subsets of $n$ points sampled from $U[0,1]^d$. However,
Arias-Castro et al.~(2005) show, in the regression setting with
 $n \times \ldots \times n=n^d$ observations on a regular $d$-dimensional grid, 
that for many relevant classes of scanning windows, such as axis-parallel rectangles
and balls, the `effective dimensionality' of the multiple testing problem is essentially linear
in the sample size $N=n^d$. That is, by showing that many of these scanning windows have a 
sizable overlap and are therefore highly correlated,
they show that the scan (\ref{scan2}) over all $\sim n^{2d}$ windows behaves like the maximum
of only $\sim N=n^d$ independent normals. Walther~(2010) gives a corresponding  construction
for $N=n$ random design points in $\R^d$, which likewise shows that scanning over all $\sim n^{2d}$
axis-parallel rectangles that contain distinct subsets of the sample reduces to a problem
that is essentially linear in the sample size $N=n$ (so the reduction in the random design
case is formally more pronounced than in the case of a regular grid with $N=n^d$ observations).
 Arias-Castro et al.~(2005)  derive the detection threshold
$\mu = \mu (N,w) \geq (\sqrt{2} +\eta) \sqrt{\frac{\log N}{w}}$
for any fixed $\eta > 0$, i.e. the exponent of effective dimension is at most 1.
 Walther~(2010) shows that with the blocked scan
it is in fact possible to attain the better threshold
(\ref{optimal}), i.e. one can replace
$\log N$ in the above threshold by $\log \frac{N}{w}$. This means that for window sizes of, say,
$\frac{w}{N} =N^{-\frac{1}{2}}$, the exponent of effective dimension is in fact $\frac{1}{2}$.
Datta and Sen~(2018) arrive at this result with the calibration~(\ref{DS}).

A further advantage of using scale-dependent calibrations is
that these calibrations allow to remove the `curse of dimensionality':
if the signal is supported on a lower-dimensional marginal, then the d-dimensional
scan with a scale-dependent calibration has essentially the same detection power as an optimal lower-dimensional  
scan would have, so there is no fundamental additional price to pay for scanning in the higher-dimensional space,
see Walther~(2010) for details.

Other related results in the multivariate setting are given by
Arias-Castro et al. (2011), who consider optimal detection when scanning over a network, while
Datta and Sen (2018) and K\"{o}nig et al. (2018) derive related results when the error distribution
is a Brownian sheet on $[0,1]^d$ and an exponential family on a rectangular grid, respectively.
Sharpnack and Arias-Castro (2016) derive the asymptotic null distributions for various
calibrations of the scan over a d-dimensional grid.

As pointed out in Section~\ref{finiteopt}, asymptotic optimality properties are typically not sufficient
to ascertain a good performance of scanning methodology. This applies especially in the multivariate setting, 
where the massive multiple
testing inherent in multivariate scanning may lead to a disappointing performance that is not adequately
reflected by asymptotic performance guarantees, even for a large sample size. It is therefore advisable
to evaluate and compare procedures with finite sample criteria, such as the {\sl realized exponent}
mentioned in Section~\ref{finiteopt}, which measures how close the procedure comes to attaining the
exponent of effective dimension described above. That is, the realized exponent is the finite
sample counterpart to the asymptotic concept of the exponent of effective dimension, and it reflects
the actual performance of a procedure in a given setting with a given sample size.

\section{Computing the scan statistic} \label{computation}

Evaluating the scan statistic, whether in its traditional form (\ref{scan}) or with any of the calibrations
discussed above, requires to evaluate $\sum_{i=1}^k Y_i$ for $1 \leq j <k \leq n$. Thus a straightforward
implementation results in a $O(n^2)$ algorithm, which makes the computation infeasible even for problems
of moderate size. However, it turns out that it is possible to approximate the scan with sufficient
precision using algorithms that have linear complexity, up to logarithmic terms.
The key idea for a fast but accurate approximation is the observation that if one examines the
interval $[j,k]$ where $j$ and $k$ are far apart, then not much is gained by also examining
intervals with similar endpoints, such as $[j-1, k+1]$. 
Therefore, when examining large intervals, it suffices to consider endpoints
on a coarse grid: If the grid spacing is small relative to the length of the interval, then
the approximation error will be small. Shorter intervals require a finer grid spacing 
for a good approximation. This idea can be used to construct a collection of intervals that is rich enough
to allow optimal detection (i.e. has small enough approximation error), while it is also sparse
enough to allow fast computation (i.e. there should be no more than about $O(n \log n)$ intervals). 
The heuristic reason why this is possible is that the $O(n^2)$ complexity is largely due to the larger
intervals, and it is precisely those that can be approximated well with a coarse grid, thus affording
considerable savings.

In more detail, such a construction can be implemented with a dyadic decomposition of the interval length, so 
that intervals of roughly the same length (up to a factor of 2, or some other factor) will be evaluated
with a certain grid spacing: At level $\ell$, we consider intervals $(j,k]$ with lengths
$k-j \in [2^{\ell},2^{\ell+1})$ and endpoints $j,k$ on a grid with grid spacing $d_{\ell}=\lceil 2^{\ell}/
\sqrt{2 \log (en2^{-\ell})}\rceil$, see Walther~(2010) for the general multivariate
construction and Walther and Perry~(2019) for the special case of the univariate regression setting (\ref{model}). 
Similar approximation schemes have been used successfully in numerical analysis and related
areas, see in particular Neill and Moore~(2004), Arias-Castro et al.~(2005) and Sharpnack and
Arias-Castro~(2016) in the context of scanning.
However, the key feature of the above construction is the particular choice of the grid spacing
$d_{\ell}$. Letting the grid spacing grow with this particular rate is important for achieving the
detection boundary (\ref{optimal}) for all window widths $w$, while still allowing efficient computation.
Moreover, in the case of the Bonferroni scan it is imperative to choose the grid spacing $d_{\ell}$
in just the right way, as too many intervals would result in too large a Bonferroni correction that diminishes
detection power. It is remarkable that such a construction exists which is both rich enough for optimal
detection and also sparse enough for efficient computation as well as for achieving optimal detection with
a simple Bonferroni correction. Thus this construction is closely connected to  fundamental aspects of
the theory behind scanning.

Fast computation with $O(n \log n)$ algorithms makes it possible to obtain critical values by
simulation rather than by analytical derivation of the distribution of maxima such as (\ref{scan}), which
requires sophisticated theoretical work, see e.g. Siegmund~(1986), Loader~(1991), 
Siegmund and Venkatraman~(1995), Kabluchko~(2011) 
or Sharpnack and Arias-Castro~(2016). In particular, the use of size-dependent critical values has
been greatly facilitated by the availability of these fast algorithms.

Finally, the above results extend to the multivariate setting. That is, there are suitable approximating sets
for the collection of all axis-parallel rectangles (or other common geometric objects) that have a cardinality
that is linear in the sample size, up to logarithmic terms. This was shown for the regression setting
with $N=n^d$ observations on a $n \times \ldots \times n$ lattice in $\R^d$ by Arias-Castro et al.~(2005), and for
the density setting with $N=n$ observations by Walther~(2010). The heuristic behind this result is that, just as
$\frac{N}{w}$ disjoint intervals of length $w$ are required to cover the univariate lattice $\{1,\ldots, N\}$,
one needs $\frac{N}{w}$ disjoint squares of area $w$ to cover a two-dimensional lattice of area $N$. Moreover, allowing
axis-parallel rectangles rather than squares will result only in additional logarithmic factors: A 
square with side length $a$ can be used as a basis to approximate rectangles with the same area and the same left endpoint 
by considering the set of rectangles with sides $m^k a$ and $m^{-k}a$ for $k=0,\pm 1,\pm 2,\ldots$ and some $m>1$. 
Due to the geometric progression of the aspect ratio, a logarithmic number of indices $k$ will suffice for a
good approximation, just as the simultaneous consideration of all windows sizes $w$ resulted in only a logarithmic
factor due to dyadic decomposition of the window sizes.

\subsection*{References}

\begin{description}
\item[] Arias-Castro, E., Cand\`{e}s, E. J., and Durand, A. (2011). Detection of an anomalous
cluster in a network. {\sl Ann. Statist.} {\bf 39}, 278--304.
\item[] Arias-Castro, E., Castro, R.M., T\'{a}nczos, E. and Wang, M. (2018). Distribution-free
detection of structured anomalies: Permutation and rank-based scans.
{\sl J. Amer. Statist. Assoc.} {\bf113}, 789--801.
\item[] Arias-Castro, E., Donoho, D. and Huo, X. (2005). Near-optimal detection of geometric
objects by fast multiscale methods. {\sl IEEE Trans. Inform. Theory} {\bf 51}, 2402--2425.
\item[] Chan, H.P. (2009). Detection of spatial clustering with average likelihood ratio test 
statistics. {\sl Ann. Statist.} {\bf 37}, 3985–4010.
\item[] Chan, H.P. and Walther, G. (2013). Detection with the scan and the average likelihood
ratio. {\sl Statistica Sinica} {\bf 23}, 409--428.
\item[] Cressie, N. (1977). On some properties of the scan statistic on the circle and the line.
{\sl J. Appl. Prob.} {\bf 14}, 272--283.
\item[] Datta, P. and Sen, B. (2018). Optimal inference with a multidimensional multiscale statistic.
arXiv:1806.02194
\item[] D\"{u}mbgen, L. and Spokoiny, V.G. (2001).  Multiscale testing of qualitative hypotheses.
{\sl Ann. Statist.} {\bf 29}, 124–152.
\item[] D\"{u}mbgen, L. and Walther, G.~(2008). Multiscale inference about a density.
{\sl Ann. Statist.} {\bf 36}, 1758–1785.
\item[] Frick, K., Munk, A. and Sieling, H. (2014). Multiscale change point inference.
{\sl J. R. Stat. Soc. Ser. B.} {\bf 76}, 495--580.
\item[] Glaz, J., Naus, J. and Wallenstein, S. {\sl Scan Statistics.} Springer Series in Statistics.
Springer-Verlag, New York, 2001. 
\item[] Kabluchko, Z. (2011). Extremes of the standardized gaussian noise. {\sl Stochastic Processes 
and their Applications} {\bf  121}, 515--533. 
\item[] K\"{o}nig, C., Munk, A. and Werner, F. (2018). Multidimensional
multiscale scanning in exponential families: Limit theory and statistical consequences. arXiv:1802.07995
\item[] Kulldorff, M. (1997). A spatial scan statistic. {\sl Commun. Statist. Th. Meth.} {\bf 26}, 
1481--1496.
\item[] Leadbetter, M. R., Lindgren, G. and Rootz\'{e}n, H. {\sl Extremes and related properties 
of random sequences and processes.} Springer Series in Statistics. Springer-Verlag, New York, 1983.
\item[] Loader, C. R. (1991). Large-deviation approximations to the distribution of the scan statistics.
{\sl Adv. Appl. Prob.} {\bf 23}, 751--771.
\item[] Nagarwalla, N. (1996). A scan statistic with a variable window.
{\sl Statistics in Medicine} {\bf 15}, 845--850.
\item[] Naus, J. I. and Wallenstein, S. (2004). Multiple window and cluster size scan procedures.
{\sl Meth. Comp. Appl. Probab.} {\bf 6}, 389--400.
\item[] Neill, D. and Moore, A. (2004). A fast multi-resolution method for detection of significant
spatial disease clusters. {\sl Adv. Neur. Info. Proc. Sys.} {\bf 10}, 651–658.
\item[] Neill, D.B. (2009). An empirical comparison of spatial scan statistics for outbreak detection.Internat. 
{\sl Journal of Health Geographics} {\bf 8}, 1–16.
\item[] Rivera,  C. and Walther, G. (2013). Optimal detection of a jump in the intensity of a Poisson
process or in a density with likelihood ratio statistics. {\sl Scand. J. Stat.} {\bf 40}, 752-769.
\item[] Rufibach, K. and Walther, G. (2010). The block criterion for multiscale inference about
a density, with applications to other multiscale problems. {\sl J. Comp. Graph. Statist.} {\bf 19},
175--190.
\item[] Sharpnack, J, and Arias-Castro, E. (2016). Exact asymptotics for the scan statistic and fast
alternatives. {\sl Elect. J. Statist.} {\bf 10}, 2641-2684.
\item[] Siegmund, D. (1986). Boundary crossing probabilities and statistical applications.
{\sl Ann. Statist.} {\bf 14}, 361--404.
\item[] Siegmund, D. and Venkatraman, E.S. (1995). Using the generalized
likelihood ratio statistic for sequential detection of a change-point. {\sl Ann.
Statist.} {\bf 23}, 255--271.
\item[] Walther, G. (2010). Optimal and fast detection of spatial clusters with scan statistics.
{\sl Ann. Statist.} {\bf 38}, 1010--1033.
\item[] Walther, G. and Perry, A. (2019). Calibrating the scan statistic: finite sample performance vs.
asymptotics. arXiv:2008.06136
\item[] Walther, G. (2021). Tail bounds for empirically standardized sums. arXiv:2109.06371
\end{description}

\end{document}